\documentclass{report}
\usepackage[dvips]{graphicx}

\begin{document}

\centerline{Dynamics of Enceladus and Dione inside the 2:1 Mean-Motion Resonance under Tidal Dissipation}
\vspace{0.5cm}\centerline{N. Callegari
Jr. and T. Yokoyama} \centerline{\small{\emph{Departamento de
Estat\'{\i}stica, Matem\'{a}tica Aplicada e Computa\c {c}\~{a}o,
UNESP, Rio Claro/SP/Brasil}}}

\vspace{5cm}

\begin{abstract}

In a previous work (Callegari and Yokoyama 2007, Celest. Mech. Dyn. Astr. vol.
98), the main features of the motion of the pair Enceladus-Dione were
analyzed in the frozen regime, i.e., without considering the tidal
evolution. Here, the results of a great deal of numerical simulations of a pair of
satellites similar to Enceladus and Dione crossing the 2:1 mean-motion
resonance are shown. The resonance crossing is modeled with a linear tidal theory,
considering a two-degrees-of-freedom model written in the
framework of the general three-body planar problem. The main regimes of motion of the system during the passage
through resonance are studied in detail. We discuss our results comparing them with classical scenarios of
tidal evolution of the system. We show new scenarios of evolution of the Enceladus-Dione
system through resonance not shown in previous approaches of the problem.

\end{abstract}

\section*{1. Introduction}
Due to the proximity of the 2:1 mean-motion resonance, the pair En\-ce\-la\-dus-Dione (hereafter denoted by E-D) is currently trapped in a libration state where the line of conjunctions between the satellites oscillates around the pericenter of Enceladus with period of 11.19 years and amplitude of $1.505^{\circ}$ (Sinclair 1972, 1983; Christou et al. 2007). The current resonant orbital configuration of the Saturnian satellites is not primordial, and probably was reached during the tidal-induced migration of the satellites due to forces acting on the planet and on the satellites (Goldreich 1965; Sinclair 1972, 1974; Sinclair 1983; Henrard and Lemaitre 1983; Peale 1986, 1999, 2003; Meyer and Wisdom 2008). The formation of the current configuration of the system is not well understood yet in spite of much effort done in the works mentioned above.

Several authors (e.g. Sinclair 1972; Henrard and Lemaitre 1983; Peale 1986) have studied the evolution of the E-D system through resonance with simplified versions of the general three-body problem given by one-degree-of-freedom models. The justification to use simplified models of resonance is based on the fact that the multiplet of resonances at 2:1 commensurability are well separated since Saturn's oblateness is large (see discussion in Peale 1986). Some works (Ferraz-Mello and Dvorak 1987; Callegari and Yokoyama 2007 ---hereafter denoted by CY2007; Meyer and Wisdom 2008), explore the dynamics of a pair of satellites similar to E-D with more general models of the three-body problem. In CY2007, a two-degrees-of-freedom model is developed to study in details the loci and the domain of the main and secondary resonances inside the 2:1 resonance. Only the conservative dynamics of the system in different portions of the phase space is studied in CY2007. Meyer and Wisdom (2008) investigate the past and future evolution of the pair Enceladus-Dione with a general model of the 2:1 resonance which also includes tidal dissipation on Saturn, Enceladus and Dione.

Here, the non-conservative dynamics of Enceladus and Dione during the passage through 2:1 resonance is investigated with an extension of the previous model (CY2007), including now the time variations of the coefficients of the model due to tides on Saturn. We show the results of evolution through resonance considering initial parameters in a wide portion of the phase space. Our methodology is similar to that developed by Tittemore and Wisdom (1988, 1990) (hereafter denoted by TW1988, TW1989, TW1990, respectively) and Peale (1988, 1999). These authors investigated the dynamics of several resonances amongst the Uranian satellites. In particular, Peale (1988), TW1990 and Peale (1999) discussed the dynamics of the pair Ariel and Umbriel through the 2:1 resonance. TW1990 estimated the probabilities of capture and escape from the 2:1 resonance in a wide range of initial values of eccentricities. Here we follow the methodology of TW1990 and apply it to the case of Enceladus and Dione.

This paper has two main purposes: 1) to show a \emph{detailed} description of the dynamics of a pair of regular satellites crossing the 2:1 mean-motion resonance in the planar case under the action of oblateness and perturbations of the central body; 2) to discuss the possibility to recover the past history of the pair E-D with our two-degrees-of-freedom model.

We have divided the work as follows. The model and the initial parameters are given briefly in Section 2. The results of a large deal of numerical simulations are shown in Section 3, where a classification of the types of the motion during the passage of the resonance is also given. In Section 4, we discuss the results obtained in Section 3 comparing them with those obtained with integrable models and most recent works. Section 5 is devoted to general conclusions.

\section*{2. Modeling the passage through the 2:1 resonance}
\subsection*{2.1 Model}
In this work we use the averaged-planar model developed in CY2007. The model is based on the classical Laplacian expansions of disturbing function, and is constructed in the
framework of the general three-body problem. The secular effects due to the perturbations of the non-spherical shape of Saturn are included considering $J_2$ and $J_4$ oblateness terms. Only the main resonant and secular terms are kept in the expression of the disturbing function.

Denoting by $m_i$, $a_i$, $n_i$, $e_i$, $\lambda_i$, $\varpi_i$ the values of mass, semi-major axis, mean motion, eccentricity, mean longitude and the longitude of pericenter of the Enceladus (subscript E) and Dione (subscript D), respectively, we consider the following set of resonant variables
\begin{eqnarray}
I_E &=&L_E-G_E,\hspace{1cm} \hspace{0.5in}\sigma_E=2\lambda_D-\lambda_E-\varpi_E; \nonumber \\
I_D &=&L_D-G_D,\hspace{1cm} \hspace{0.5in}\sigma_D=2\lambda_D-\lambda_E-\varpi_D; \nonumber \\
\Gamma_E &=&L_E+(I_E+I_D),\hspace{1.6cm}
\hspace{0.35in}\lambda_E; \label{1}\\
\Gamma_D &=&L_D-2(I_E+I_D),\hspace{1.4cm}\hspace{0.35in}%
\lambda_D, \nonumber
\end{eqnarray}
where $L_i=\beta_i\sqrt{\mu_i a_i}$, $G_i=L_i\sqrt{1-e^2_i}$,
$\beta_i=\frac{Mm_i}{M+m_i}$, $\mu_i=$G$(M+m_i)$, $M$ is the Saturn mass, G=1293.4686 is the gravitational constant in units of day, equatorial radius of Saturn ($R_e$) and $M$, whose values are listed in Table I.

In terms of the non-singular variables $x_i=\sqrt{2I_i}\cos \sigma_i$, $y_i=\sqrt{2I_i}\sin \sigma_i$, and neglecting the constant terms, the expanded Hamiltonian can be written as
\begin{eqnarray}
H &=&A(x_E^2+y_E^2+x_D^2+y_D^2)\nonumber \\
&&+B(x_E^2+y_E^2+x_D^2+y_D^2)^2
\nonumber \\
&&+C(x_E^2+y_E^2)+D(x_D^2+y_D^2)+E(x_Ex_{D}+y_Ey_{D})
\label{2}\\
&&+Fx_E+Ix_D \nonumber \\
&&+K(x_Ex_D-y_Ey_D)+R(x_E^2-y_E^2)+S(x_D^2-y_D^2)\nonumber,
\end{eqnarray}
where $A-F$, $I$, $K$, $R$ and $S$ are the coefficients of the Hamiltonian which depend on the masses, $J_2$, $J_4$, eccentricities, semi-major axes, and the ratio $\alpha=\frac{a_E}{a_D}$ through combinations of the Laplace coefficients. The expressions and numerical values of the coefficients are given in Table II (first and third columns). They are calculated in the exact 2:1 commensurability (see Table I). In terms of resonant variables defined by Eq. (\ref{1}), the coefficients are multiplied by the quantities given in the second column in Table II. Note that since the averaged Hamiltonian (\ref{2}) is cyclic in $\lambda_E$ and $\lambda_D$, $\Gamma_E$ and $\Gamma_D$ are constants of motion, and the dynamical system has two degrees of freedom.

\newpage

\textbf{\emph{Table I:} Model parameters and tidal constants.}

\centerline{
\begin{tabular}{ccccc}
\hline            & Saturn                    & Enceladus               & Dione \\
\hline
$J_2$, $J_4$$^a$  & 0.0162907, -0.00093583    &    -                    &    -  \\
Mass$^a$          & $3498.96^{-1}$ $M_{\odot}$   &$1.901 \times 10^{-7}M$  &$19.275\times 10^{-7}M$\\
Radius (km)$^{a,b}$               & $60268$   &$252.3$                  &$562.5$\\
$Q$: dissipation function$^c$     &$>$18,000  &$100^d$                  &$100^{e}$ \\
$k_2$: dynamical Love number$^c$  &0.34       &$0.00075^d$              &$0.00075^e$\\
Eccentricity                      & -         &$0$                      &$0$\\
Semi-major axes in the exact      & -         &$3.938806$               &$6.252469$\\
2:1 commensurability ($R_e$)      &           &                         &          \\
Initial semi-major axis$^f$       & -         &$3.938579$               &$6.252288$\\
Current semi-major axis$^b$       & -         &$3.955812$               &$6.266127$\\
Current eccentricity$^{b,g}$      & -         &$0.0045$                 &$0.0022$\\
\hline
\end{tabular}}
{\small $^a$ Jacobson et al. (2006). $M_{\odot}$ is the solar mass.}

{\small $^b$ http://ssd.jpl.nasa.gov/horizons.html, January 2007.}

{\small $^c$ Murray and Dermott (1999).}

{\small $^d$ In Meyer and Wisdom (2008), $Q_E=20$ and $k_{2E}=0.0018$ are also adopted.}

{\small $^e$ Supposing the same value as the Enceladus one.}

{\small $^f$ Values considered in the calculations of the initial $\delta$ parameter. }

{\small $^g$ See also Sinclair (1972), Ferraz-Mello (1985).}

\newpage

\textbf{\emph{Table II:} Expressions of the coefficients of the Hamiltonian and integrals $\Gamma_i$ and their numerical values.}

\centerline{
\begin{tabular}{ccccc}
\hline Coefficient&Terms in Hamiltonian & Numerical Value \\
\hline
$A=\frac{1}{2}GM\left[ -\frac{k_E}{\Gamma_E^3}+\frac{2k_D}{\Gamma_D^3}\right]$&$2(I_E+I_D)$&$-5.2838\times10^{-7}$ \\
$B=-\frac{3}{8}GM\left[ \frac{k_E}{\Gamma_E^4}+\frac{4k_D}{\Gamma_D^{4}}\right]$ &$4(I_E+I_D)^2$&$-147057.5727$ \\
$C=\frac{1}{2}k_{12}\left\{ \frac{-b}{2\Gamma_E\Gamma_D^2}+\frac{4a}{\Gamma_D^3}+\frac{2}{\Gamma_D^2}\frac{da}{d\alpha }\frac{\beta_D^2\mu_D%
}{\beta_E^2\mu_E}\left[ \frac{\Gamma_E}{\Gamma_D^2}+\frac{2\Gamma_E^2}{\Gamma_D^3}\right]\right\}+T_{j2c}+T_{j4c}$ &$2I_E$&$-8.043543\times10^{-3}$\\
$D=\frac{1}{2}k_{12}\left\{ \frac{1}{\Gamma_D^3}\left[ -\frac{b}{2}+4a\right] +%
\frac{2}{\Gamma_D^2}\frac{da}{d\alpha }\frac{\beta
_D^2\mu_D}{\beta_E^2\mu_E}\left[ \frac{\Gamma_E}{\Gamma_D^{2}}+\frac{2\Gamma_E^2}{%
\Gamma_D^3}\right]\right\}+T_{j2d}+T_{j4d}$  &$2I_D$&$-5.1166716\times10^{-3}$\\
$E=-\frac{ck_{12}}{4\Gamma_D^{2}\sqrt{\Gamma_E\Gamma_D}}$&$2\sqrt{I_EI_D}\cos(\sigma_E-\sigma_D)$&$ 8.998116\times10^{-7}$\\
$F=-\frac{dk_{12}}{2\Gamma_D^{2}\sqrt{\Gamma_E}}$&$\sqrt{2I_E}\cos \sigma_E$&$2.4498297\times10^{-8}$ \\
$I=-\frac{ek_{12}}{2\Gamma_D^{2}\sqrt{\Gamma_D}}$&$\sqrt{2I_D}\cos \sigma_D$&$-6.0934146\times10^{-9}$\\
$K=-\frac{hk_{12}}{4\Gamma_D^{2}\sqrt{\Gamma_E\Gamma_D}}$&$2\sqrt{I_EI_D}\cos(\sigma_E+\sigma_D)$&$7.7631926\times10^{-6}$ \\
$R=-\frac{fk_{12}}{8\Gamma_E\Gamma_D^2}$&$2I_E\cos2\sigma_E$ &$-5.353961\times10^{-6}$\\
$S=-\frac{gk_{12}}{8\Gamma_D^3}$& $2I_D\cos2\sigma_D$&$-1.571576\times10^{-6}$\\
$\Gamma_E$& &$1.3568827\times10^{-5}$ \\
$\Gamma_D$& &$1.7333956\times10^{-4}.$ \\
\hline
\end{tabular}}
In Table 2, $a-h$ are combinations of Laplace coefficients (e.g. Murray and Dermott 1999), and \begin{eqnarray}&&k_E=\mu_E\beta_E^2m_E,
k_D=\mu_D\beta_D^2m_D, k_{12}=Gm_Em_D\mu_D\beta_D^2;\nonumber\\
&&T_{j2c}=4.5C_1-6C_2, T_{j4c}=8C_3-10C_4, T_{j2d}=3C_1-4.5C_2, T_{j4d}=5C_3-28C_4;\nonumber\\
&&C_{1}=\frac{c_{11}\beta_E^6\mu_E^3}{\Gamma_E^7},\hspace{0.3cm}
C_2=\frac{c_{12}\beta_D^{6}\mu_D^3}{\Gamma_D^7},\hspace{0.3cm}
C_3=\frac{c_{21}\beta_E^{10}\mu_E^5}{\Gamma_E^{11}},\hspace{0.3cm}
C_4=\frac{c_{22}\beta_D^{10}\mu_D^5}{\Gamma_D^{11}};\nonumber\\
&&c_{11}=-\frac{1}{2}GMm_ER_{e}^2J_2,
c_{12}=-\frac{1}{2}GMm_DR_e^2J_2,
c_{21}=+\frac{3}{8}GMm_ER_e^4J_4,
c_{22}=+\frac{3}{8}GMm_DR_e^4J_4\nonumber.
\end{eqnarray}

\subsection*{2.2 $\delta$ Parameter. Tidal parameters and the rate of evolution through resonance}

The passage of the pair E-D through the 2:1 resonance is modeled here by including the time variation of the Hamiltonian's coefficients, following the same strategy given in (TW1988; TW1990). Due to
tidal dissipative processes, the semi-major axes and eccentricities of the satellites suffer slow variations. Among all
coefficients in Eq. (\ref{2}), the most sensitive one to such variations is $A$. In fact, expanding the expression of $A$ in a double series of
$I_i$ up to first order, it can be shown that, in the proximity of resonance and for near-circular orbits,

\begin{equation}
A\approx\frac{1}{2}(2n_D-n_E)-4B(I_E+I_D).\label{4}
\end{equation}
The first member at right-hand side of Eq. (\ref{4}) corresponds to the resonant combination of the mean motions which, in the proximity of the exact 2:1 commensurability, suffers large variations. All other coefficients of the Hamiltonian will be kept constant because they suffer small variations in the vicinity of the resonance.

In order to study the system considering a wide variety of initial configurations in the vicinity of the exact resonance, we define the parameter

\begin{equation}
\delta \equiv 4A + 2(C+D).\label{5}
\end{equation}
In a first approximation and for small eccentricities, $\delta\approx4n_D-2n_E-\dot\varpi_E-\dot\varpi_D$.
The value of $\delta$ changes sign in the middle of the exact resonance when we consider only the resonant term factored by $K$ in Hamiltonian (\ref{2}).

Adopting $\delta$ as a free parameter, we can model the passage of our dynamical systems through the resonance. The rate of variation of $\delta$ is given by

\begin{eqnarray}
&&\dot\delta\approx 4\dot A\label{55}\\
&&\dot A\approx \frac{1}{2}(2\dot n_D-\dot n_E)-4B(\dot I_E+\dot I_D)\label{6}\\
&&\dot I_i\approx I_i\left(-\frac{1}{3}\frac{\dot n_i}{n_i}+2\frac{\dot e_i}{e_i}\right)\label{7}
\end{eqnarray}

In Eq. (\ref{55}), the variations of the coefficients C and D have been neglected since their values ($\sim2\times10^{-16}$) are hundreds of times smaller than $\dot A$. In order to evaluate $\dot A$, we substitute the \emph{current} values of the parameters of the system given in Table I in the following expressions:

\begin{eqnarray}
&&\dot n_i=+\frac{63}{2}\frac{n^2_ik_{2i}M}{Q_im_i}\left(\frac{R_i}{a_i}\right)^5e^2_i -\frac{9}{2}  \frac{k_{2S}}{Q_S} \frac{m_i}{M}\left(\frac{R_e}{a_i}\right)^5 n^2_i\left(1+\frac{51}{4}e^2_i\right),
\label{8}\\
&&\dot
e_i=-\frac{21}{2}\frac{k_{2i}}{Q_i}\frac{M}{m_i}\left(\frac{R_i}{a_i}\right)^5n_ie_i+
\frac{57}{8}\frac{k_{2S}}{Q_S}\frac{m_i}{M}\left(\frac{R_e}{a_i}\right)^5n_ie_i,\label{9}
\end{eqnarray}
where, as usual, $k_{2i}$, $Q_i$, $R_i$ refer to Love number, tidal dissipation function and mean radii of the satellites, respectively; the subscript $S$ is used for Saturn (see Table I). Eqs. (\ref{8}, \ref{9}) give us the time variation of the mean orbital elements of a satellite in a tidal model where the following approximations and hypothesis have been assumed (Ferraz-Mello et al. 2008): i) All dissipation functions are independent of the frequency of the corresponding tidal wave; ii) the satellite is in synchronous or in super-synchronous rotational state; iii) the satellite's mean motion is smaller than the frequency of the planet rotation; iv) the tidal potential is truncated in second order of eccentricities, and the inclination terms have been neglected. The first and second terms on the right in Eqs. (\ref{8}, \ref{9}) are due to tides raised on the satellite by the planet, and tides raised on the planet by the satellite, respectively.

The values of $k_{2i}$ and $Q_i$ are unknown, and only some estimative of the domain of their plausible values are available. For instance, for Saturn, $Q_S\approx18,000$ is the minimum possible value, which is obtained from the current configuration of Mimas (e.g. Murray and Dermott 1999). As the Enceladus-Dione pair is currently in 2:1 resonance, we can roughly estimate the maximum value of $Q_S$: if we consider that the pair E-D crossed the 2:1 resonance in the beginning of the solar system, then we find that the corresponding $Q_S$ must be less than $350,000$; otherwise, the 2:1 resonance could not be reached during the age of the solar system. This estimative is based on the linear tide theory neglecting the terms of dissipation on the satellites in Eq. (\ref{8}), and taking into account only E-D pair. Of course, a better estimative must be in close agreement with other resonant pair of the Saturn's system (Titan-Hyperion, Mimas-Tethys etc). For instance, assuming that the pair Mimas-Dione has crossed the 3:1 mean-motion resonance, $Q_S$ is given in the interval $[18,000;100,000]$ (Meyer and Wisdom 2008).

In our numerical experiments we have used the intermediary value of $Q_S=34,000$. This value is slightly larger than that given by Ferraz-Mello and Hussmann (2005). Adopting $Q_S=34,000$, our numerical experiments show a very smooth evolution of the pair of satellites through the resonance, a condition which must be satisfied in order to avoid unrealistic evolutionary scenarios (TW1988).

Fig. 1(a) shows $\dot A$ as functions of $Q_S$. In our simulations, we have used $\dot A\approx4.4\times10^{-14}$, corresponding to the vertical full line in Fig. 1(a) ($Q_S=34,000$). The dashed line in Fig. 1(a) indicates the larger value for $\dot A$, which is about $8.2\times10^{-14}$.

\subsection*{2.3 Enceladus' Equilibrium eccentricity}

Consider a planet and one satellite where Eqs. (\ref{8}, \ref{9}) hold. Following TW1990, we calculate the non-resonant equilibrium eccentricity, for which the dissipation due to the tides on a synchronous or stationary rotating satellite ceases the growth of the semi-major axis due to planetary tides:

\begin{eqnarray}
e_{eq.}\approx\sqrt{\frac{1}{7D-\frac{51}{4}}}, \hspace{0.6cm}D=\frac{k_{2i}}{k_{2S}}
\frac{Q_S}{Q_i}
{\left(\frac{M}{m_i}\right)}^2{\left(\frac{R_i}{R_e}\right)}^5\label{11}.
\end{eqnarray}
Eq. (\ref{11}) is obtained by setting $\dot n=0$ in Eq. (\ref{8}). In order to evolve through resonance, the satellite initial eccentricity must be less than the equilibrium eccentricity. The equilibrium eccentricity depends on the Love number and dissipation function of the satellites and the planet. In Fig. 1(b) we plot $e_{eq.}$ of Enceladus as a function of $k_{2E}$ for some values of $Q_S$, $Q_E$. The two vertical lines at $k_{2E}\approx0.0007$ and $\approx0.0018$ refer to two values of the Love number for Enceladus found in literature (see Table I). No matter $Q_S$, $Q_E$ and $k_{2E}$ we adopt in Fig. 1(b), the important point here is that equilibrium eccentricity is $e_{eq.}\geq0.02$. Therefore we kept the initial value of the Enceladus eccentricity below the critical value $0.02$ in almost all numerical experiments shown in Section 3.

\subsection*{2.4 Approaching the resonance zone}
The dynamics of the system (\ref{2}) depends strongly on the parameters (semi-major axes, masses, tidal constants), and on the initial conditions (eccentricities and initial critical angles). Following TW1988, we study the passage through the 2:1 resonance considering different initial values of eccentricities ($e_{E0}$, $e_{D0}$), keeping in all cases the same initial values of the semi-major axes (Table I).

For each pair of initial eccentricities, we generate sets of initial conditions in the following way: without considering tide and adopting initially $\sigma_E=\sigma_D=0$, Hamiltonian (\ref{2}) is integrated over a short time span (about some years). Then we select on this orbit 20 different points, all of them having the same energy and slightly different eccentricities, but distinct values of $\sigma_E$ and $\sigma_D$. Each of these points will be used as the initial condition to be integrated for a long time, but now the tidal effect will be considered through the integration of $\dot \delta$. Each set with 20 numerical simulation will be denoted by run 1, 2, 3 etc in next section.

\section*{3. Numerical Simulations}
In this section we describe the main regimes of motion of the pair E-D around the resonance considering different ranges of initial conditions, according to the methodology given in Section 2.4. After a detailed analysis of the main regimes of motion, we also estimate the probability of captures into some regimes of motion. For these tasks, we investigate the values of the eccentricities ($e_{E}$, $e_{D}$) and critical angles ($\sigma_E$, $\sigma_D$) reached by the orbits during the evolution through resonance. In order to better visualize the results, the plots of the variables are given in function of the $\delta$ parameter. Let the total length of the time variation of $\delta$ be divided in successive intervals $\Delta\delta$ (typically $\Delta\delta\approx10^{-5}$ or less). For each interval, in general we plot only the maximum and minimum value of the eccentricities and critical angles.

Very far from the resonance encounter ($\delta\ll0$), the long-term evolution of the pair of satellites is dominated by secular interactions, where the mean eccentricities oscillate regularly and the critical angles $\sigma_E$ and $\sigma_D$ circulate (TW1988; Callegari et al. 2006). As the system approaches the resonance, different scenarios of capture and escape from resonance may occur.

\subsection*{3.1 Scenario for initial eccentricities smaller than $0.0035$}

Let us begin considering values of initial eccentricities in two subintervals: $0\leq e_{E0}\approx e_{D0}\leq0.0015$ and $0.0015\leq e_{E0}\approx e_{D0}\leq0.0035$. The main characteristics of the dynamics of the system inside the resonance in these intervals of initial eccentricity are summarized in Figs. 2(a-h).

For $0\leq e_{E0}\approx e_{D0}\leq0.0015$, $\sigma_E$ is always captured into libration about zero in the beginning of the simulation. Fig. 2(c) shows an example where the capture occurs at $\delta\approx-0.015$. Fig. 2(d) shows the evolution of $\sigma_D$ corresponding to the same orbit given in Fig. 2(c). $\sigma_D$ is captured slightly later into libration about $\pi$ at $\delta\approx-0.006$. In this example, in spite of capture of $\sigma_E$, Enceladus eccentricity keeps its value near the initial one (Fig. 2(a)), as the system evolves through resonance. In the case of $\sigma_D$, its capture implies a large growth of mean eccentricity of Dione (Fig. 2(b)). Based on previous studies (CY2007), the libration of $\sigma_E$ resembles quite well the paradoxical (or small-eccentricity) libration. On the other hand, the libration of $\sigma_D$ corresponds to a true libration (regime RDI in CY2007). We denote the aforementioned motions of $\sigma_E$ and $\sigma_D$ by PL(0) (paradoxical) and L($\pi$) (true), respectively.

For $\delta\geq-0.006$, Figs. 2(c,d) also show that the libration of $\sigma_E$, $\sigma_D$ and $\Delta\varpi$ occur simultaneously. This kind of motion of the system was first described in CY2007, and may occur when the system attains values of energy near the maximum. In order to better understand this regime of motion inside resonance (i.e, simultaneous libration of the critical angles), we compute surfaces of sections at several values of $\delta$ (Fig. 3). Enceladus sections are defined through the condition $y_D=0$ (see Eq. (\ref{2})), represented on the plane ($e_E\cos \sigma_E\times e_E\sin\sigma_E$). Dione section is defined through the condition $y_E=0$, and is represented on the plane ($e_D\cos \sigma_D\times e_D\sin\sigma_D$) (CY2007). Figs. 3(a,b) show the section for $\delta=+0.007$. The energy is $H=18.2\times10^{-12}$, very near the maximum of energy ($H\approx18.22\times10^{-12}$). In this portion of the phase space, the system is in the co-rotation zone, where libration of the critical angles associated to resonance and long-period oscillation of $\Delta\varpi$ can occur at same time. Fig. 3(c) shows the time variations of the elements of a single orbit where $\Delta\varpi$ and $\sigma_D$ librate with same long-term period and amplitude. Fig. 3(d) shows the time evolution of the argument of pericenter of the orbits shown in Fig. 3(c): $\varpi_i$ are circulating with the same frequencies since $\Delta\varpi$ librates.

In the interval of initial eccentricities $0.0015\leq e_{E0}\approx e_{D0}\leq0.0035$, another type of motion often occurs: $\sigma_E$ always circulates during the evolution through resonance (Fig. 2(g)), while $\sigma_D$ is captured into libration about $\pi$ (Fig. 2(h)). In this case the conjunctions between satellites occur in a line oscillating with large amplitude around the apocenter of Dione orbit. Note that $\Delta\varpi$ circulates (Fig. 2(h)). In this example we can show that the system evolves in direction of maximum energy, but does not enter in the co-rotation zone (see Figure 11 in the Appendix).

We can summarize our results on the evolution through resonance with initial eccentricities in the interval $0\leq e_{E0}\approx e_{D0}\leq0.0035$ as follows: in all 140 numerical simulations done considering this range of initial conditions, $\sigma_D$ is \emph{always} captured into libration around $\pi$. In the subinterval $e_{i0}\leq0.0015$, the dynamics is well characterized by the example given in Figs. 2(a-d). In the case where $0.0015 \leq e_{E0}\leq0.0035$, the dominant regime of motion is given by the example shown in Figs. 2(e-h).

We finish this section discussing the run 3, where initial eccentricities are of the same order of the current values (see Table I). For these eccentricities, the dominant regime of motion in almost all simulations is similar to that given in Figs. 2(e-h). However, some exceptions may occur. For instance, Figs. 2(i-l) show a case where, after the system has entered in resonance without capture of $\sigma_E$ (similar to Figs. 2(e-h)), a sudden reduction of $e_E$ occurs at $\delta\approx0$ (Fig. 2(i)), and $\sigma_E$ is finally captured in the paradoxical libration PL(0) (Fig. 2(k)). At $\delta\approx+0.018$, the amplitude of libration increases when a jump in $e_E$ appears. The same occurs with $\sigma_E$ at $\delta\approx+0.028$, in this case, leading $\sigma_E$ to return to the circulation regime. The jumps in the curve of $e_E$ occur since the system cross the separatrix associated to some secondary resonance (see also TW1990).

Secondary resonances in the system E-D were studied in detail in CY2007, and they occur in portions of the phase space where some linear combination of $\sigma_E$ and $\sigma_D$ is commensurably. The centers of some secondary resonances in our system are illustrated in Figs. 3(e,f): the fixed points indicated by 1/1 and 1/2 correspond to the loci in the phase space where $\sigma_E+\sigma_D$ and $2\sigma_E+\sigma_D$ librate about $\pi$, respectively.

Finally we note that the behavior of $\sigma_D$ in the example given in Figs. 2(i-l) is the same as that shown in Figs. 2(b,f), except that at $\delta\approx+0.002$ the amplitude of libration increases, while it seems to cause larger oscillation of $e_D$ (Fig. 2(j)).

\subsection*{3.2 Scenario for $0.0035\leq e_{E0}\approx e_{D0}<0.017$}

Figs. 4(a-h) show two examples taken from run 8, where $e_{i0}\approx0.007$. The system crosses the resonance and $\sigma_E$ is captured into libration at $\delta\approx+0.01$. However, in opposition to the case shown in Figs. 2(a,c), the mean eccentricity of Enceladus now increases significantly during the evolution through resonance (Fig. 4(a)). This kind of motion corresponds to the capture of $\sigma_E$ into true libration about zero, which will be referred as L(0) hereafter. The fixed point associated to L(0) and the chaotic separatrix associated to resonance can be seen in surface of section shown in Fig. 3(f). The center of the true libration is located at $e_E\cos\sigma_E\approx+0.026$ for $\delta=+0.025$. Note in Fig. 3(f) the presence of several high-order secondary resonances near the separatrix of L(0). See also Figures 12(c,d) in the Appendix.

In the case of Dione, its mean eccentricity (Fig. 4(b)) suddenly decreases at $\delta\approx+0.003$, showing that the system passes through resonance without capture into libration of $\sigma_D$. Fig. 4(d) shows $\sigma_D$ circulating during the evolution through resonance.

The case shown in Figs. 4(e-h) is similar to the example discussed in Figs. 4(a-d), except that now, for $\delta\approx+0.016$, the amplitude of libration of $\sigma_E$ increases (Fig. 4(g)). Whenever $\sigma_E$ increases for a given $\delta$, we have also a corresponding effect on $e_E$ and $e_D$ (Figs. 4(e,f)). Large variation in the amplitude of the critical angle can cause the system escapes for larger values of $\delta$. See also Meyer and Wisdom (2008). As we will see in Section 3.3, escape from resonance by natural increasing of amplitude of critical angle is very often for larger values of initial eccentricities.

\subsection*{3.3 Scenario for $e_{E0}\gg e_{D0}$}

Consider now the results of numerical simulations where the initial conditions are such that $0.0035\leq e_{D0}\leq0.007$ and $e_{E0}\geq 0.017$. As we will see in Section 4, in these ranges of initial eccentricities the probability of capture into libration of $\sigma_E$ around zero is small. Figs. 4(i,k) show an example where the system passes through resonance without capture of $\sigma_E$. In the trajectory, $\sigma_D$ is temporarily captured into libration about $\pi$ at $\delta\approx 0$ (Fig. 4(l)) since the system escapes from resonance at $\delta\approx +0.016$ by increasing of the libration amplitude of $\sigma_D$ (see discussion in Section 3.2).

Figs. 5(a-d) show an example where the system crosses the resonance without any capture. In fact only a very short temporary libration around zero at $\delta\approx+0.015$ can be seen in Fig. 5(a), but the evolution of $\sigma_E$ can be considered very similar to the case shown in Fig. 4(i). In the case of Dione critical angle, $\sigma_D$ is not captured (Fig. 5(b)). On the other hand, in the trajectory shown in Figs. 5(e-h), the escape of the system from L(0) regime of motion occurs after a long period of temporary capture.

Here, let us discuss some complex evolutionary scenarios of capture and escape from secondary resonances which appear in several intervals of initial eccentricities discussed in this work. This time, none of $\sigma_E$ or $\sigma_D$ is captured into their main libration centers (Figs. 6(c,d)) but, in opposition to the cases shown in Figs. 5(a-d), both eccentricities increase (Figs. 6(a,b)). The growth in eccentricities occurs due to some capture into a secondary resonance (see Section 3.1). Figs. 6(a-d) show examples where capture into $\sigma_E+\sigma_D$ secondary resonance occurs at $\delta\approx+0.005$: the angle $\sigma_E$+$\sigma_D$ librates while $\sigma_E$ and $\sigma_D$ circulate. The scattered points at $\delta\approx+0.025$ in Figs. 6(a-d) belong to a chaotic region of the phase space which can also be identified in some surfaces of section, more precisely, in the Enceladus section (Fig. 3(f)): the source of this chaos is the separatrix associated to the true libration of $\sigma_E$ (regime L(0)).

Other examples of the growth of Enceladus eccentricity due to capture into secondary resonances are given in Figs. 6(e-l). This time, we show the libration of $\sigma_E+\sigma_D$ in Fig. 6(h) and $\sigma_E+4\sigma_D$ in Fig. 6(l). In both cases, the system escapes from secondary resonances at large values of $\delta$. This occurs due to interaction of their domain with the separatrix of true resonance L(0). Fig. 6(g) shows that, after the escape, $\sigma_E$ is briefly captured into libration and escapes. In the case shown in Fig. 6(k) we see an interesting feature: after the escape from secondary resonance, $\sigma_E$ is captured into L(0) with decreasing amplitude while $e_E$ increases. In the case of Figs. 6(i-l) it is worth to note how high order secondary resonance is able to excite the initial eccentricities of the satellites.

The role of secondary resonances in the dynamical history of natural satellites was studied by several authors. As far as we are aware, capture into secondary resonances without prior libration of the system into libration of $\sigma_E$ or $\sigma_D$, as shown in Fig. 6, seems to be new, not described in previous works (TW1989; TW1990; Malhotra and Dermott 1990; Henrard and Sato 1990).

To close this section, it is important to mention some differences when comparing the general three-body model with simplified models of resonance: in Fig. 4(i) and Figs. 5(a,b,f), the final mean eccentricity after passage through the resonance is smaller than its initial value. This is the general behavior for one-degree-of-freedom models (see Peale 1986), in the case of passage without capture. However in Fig. 4(j), Figs. 6(e,f,j) and Fig. 5(e), the final mean eccentricity of the satellite is larger than the initial one, which is not predicted by integrable models (TW1988).

\subsection*{3.4 Scenario for $e_{E0}\ll e_{D0}$}

Since the Dione mass is about 11.2 times larger than the Enceladus one (Table I), Dione's perturbation on Enceladus is stronger than the Enceladus' one over Dione. Thus, the hypothesis to consider $e_{E0}\ll e_{D0}$ may not be completely realistic unless some previous phenomenon outside the 2:1 resonance might have happened, pumping $e_D$ to higher values. In this section we only show the results for $e_{E0}\ll e_{D0}$, leaving a brief discussion on the reliability of this initial conditions to Section 4.

We already showed that quite different regimes of motion are found when we consider $e_{E0}\sim e_{D0}$ (Sections 3.1 and 3.2), and $e_{E0}\gg e_{D0}$ (Section 3.3). For $e_{E0}\ll e_{D0}$, new regimes of motion are obtained. Figs. 7(a-d) show a case where $e_{D0}\approx0.004$ and $e_{E0}\approx0.002$. At $\delta\approx-0.002$, $\sigma_E$ is temporary captured into 0 (Fig. 7(c)), and librates until $\delta\approx+0.004$, when the system escapes from resonance. After the escape, however, Enceladus and Dione eccentricities increase suddenly, since the system is captured into a secondary resonance, where the angle $2\sigma_E+\sigma_D$ librates around $\pi$ (Fig. 7(d)). At $\delta\approx+0.036$ the system escapes from secondary resonance.

The mechanism of capture into secondary resonance shown in Figs. 7(a-d) is different from those discussed in Fig. 6: there, the system was automatically captured into secondary resonances, while now the system is first captured into the main libration, escapes, and enters in a secondary resonance. This mechanism of capture into secondary resonances is similar to those reported in TW1989, TW1990, Malhotra and Dermott (1990).

Figs. 7(e-h) differ from the previous case (Fig. 7(a-d)) since $\sigma_E$ is sooner captured into libration about 0. At $\delta\approx-0.004$, the system escapes from libration (Fig. 7(g)), and $\sigma_E$ circulates until $\delta\approx+0.005$, when a new temporary capture into libration about zero occurs. The system escapes and enters into the $\sigma_E+\sigma_D$ resonance (Fig. 7(h)).

We finish Section 3 showing two examples of complex regimes of motion found only when $e_{E0}\ll e_{D0}$ and initial eccentricities are very near zero. In the case shown in Fig. 7(j), $\sigma_E$ is captured in paradoxical libration in the beginning of the simulation, but escapes and is again re-captured at $\delta\approx-0.005$. Note the similarity between Figs. 2(a) and 7(i). The case shown in Fig. 7(l) is more complicated since the system is alternating between libration and circulation during the evolution through the 2:1 resonance.

\subsection*{3.5 Statistics}

In this work a total of at least one thousand numerical simulations were performed. Although not exhaustive, we think that a significant part of the resonant domain could be covered. These simulations revealed several different dynamics, and most of them can be collected in one of the following cases:

1) Capture of $\sigma_E$ into paradoxical libration about zero (PL(0)), and capture of $\sigma_D$ about $\pi$ (L($\pi$)) (Figs. 2(a-d)). Sometimes, capture of $\sigma_E$ occurs at large values of $\delta$ and, eventually, escapes (Figs. 2(i-l)).

2) The same as 1) where $\sigma_E$ alternates between libration in paradoxical libration and circulation (Figs. 7(i-l)).

3) Passage through resonance without capture of $\sigma_E$, and capture of $\sigma_D$ into libration about $\pi$ (Figs. 2(e-f)).

4) Capture of $\sigma_E$ into true libration about zero (L(0)), and passage through resonance without capture of $\sigma_D$ (Fig. 4(a-h)).

5) Escape from temporary libration of $\sigma_E$ in L(0), and passage through resonance without capture of $\sigma_D$ (Figs. 5(e-h)).

6) Passage through resonance without capture of $\sigma_E$ and $\sigma_D$. Sometimes $\sigma_E$ quickly librates about zero (Figs. 5(a-d)).

7) Passage through resonance without capture of $\sigma_E$ into L(0), and capture of $\sigma_D$ into L($\pi$) with subsequent escape (Figs. 4(i-l)).

8) Automatic capture into secondary resonances (Fig. 6).

9) Temporary capture of $\sigma_E$ into L(0), followed by capture into secondary resonance (Figs. 7(a-d)). In some cases, the system is re-captured into L(0) at large values of $\delta$.

10) The same as 9, except that $\sigma_E$ is captured into libration in the beginning of the simulation (Figs. 7(e-h)).

11) Exceptions corresponding to cases which do not match any case listed above.

Fig. 8(a) shows that, for initial eccentricities $0\leq e_{i0}\leq0.0015$, only the case 1) is possible. For $0.0015\leq e_{i0}\leq0.0035$, $\sigma_E$ is not captured into paradoxical libration (except the three cases when $e_{i0}=0.0025$), while the capture of $\sigma_D$ about $\pi$ is certain and the most probable event is the case 3).

For initial eccentricities taken in the interval $0.0035\leq e_{i0}\leq0.016$, the capture into state 4 is a very probable event. By inspection of Fig. 8(a) we see that the probability capture of $\sigma_E$ into libration about zero increases (slightly) as long as the initial eccentricities are selected in the interval $0.005\leq e_{i0}\leq0.01$. This result is in opposition to those predicted by single resonance models, which states that capture into true libration always occurs for initial eccentricities smaller than some value $e_{max}$ (see discussion in Section 4). The cases 6 and 7 never occur for $0.0035\leq e_{i0}\leq0.016$, and the case 5 is a very rare event in this range of initial eccentricities. However, there are some cases of escape from secondary resonances (e.g. Fig. 6(e)), resulting in a slight increase in the total number of escapes for $0.0035\leq e_{i0}\leq0.016$.

Fig. 8(b) shows the statistics for $e_{E0}\gg e_{D0}$. In the two runs where $e_{E0}=0.008$, $e_{D0}=0.004$, and $e_{E0}=0.01$, $e_{D0}=0.005$, the dominant regime of motion is L(0). In all cases where $e_{E0}\geq0.017$ and $e_{D0}\geq0.0035$, the probability of passage through the resonance without any capture is high. When $e_{E0}\geq0.017$ and $e_{D0}\leq0.0035$ (as the case $e_{E0}=0.02$, $e_{D0}=0.002$), although capture of $\sigma_E$ in general is ruled out, $\sigma_D$ always falls into libration in $\pi$. Other interesting run is the case $e_{E0}=0.0045$, $e_{D0}=0.0022$ corresponding to the values of the current eccentricity: the dominant regime of motion is the case 3, and there are two exceptions where paradoxical libration appears for large values of $\delta$ (Figs. 2(i-l)).

Fig. 8(c) corresponds to the cases $e_{E0}\ll e_{D0}$. In this range of initial eccentricity, the new features are the presence of: i) capture into secondary resonances according to the cases 9 and 10; ii) several complex scenarios of capture and escape when the initial eccentricities are both very near zero (note the cases $e_{E0}=0$, $e_{D0}=0.002$ and $e_{E0}=0$, $e_{D0}=0.004$).

\subsection*{3.6 Parametric Plane}

From Eq. (\ref{5}) we write $A=\frac{1}{4}\left[\delta-2(C+D)\right]$, so that the Hamiltonian (\ref{2}) becomes a function of $\delta$. $H$ and $\delta$ are the natural parameters to describe the evolution of the system around the resonance, as pointed out in Section 2.4. In this section we show the parametric plane ($\delta,H$), which is very convenient to summarize the results about the dynamics shown in previous subsections.

The parametric plane (Fig. 9(a)) is constructed by calculating the zero-gradient points of Eq. (\ref{2}), in the case of symmetric configurations (i.e., $\sigma_E$, $\sigma_D=0,\pi$). Zero-gradient points can be seen in Fig. 9(b) in the representative plane of initial conditions, which are given by the level curves of the $H=H(e_E,e_D)$ for a fixed value of $\delta$. In Fig. 9(b) shows also, in bold lines and gray symbols, the curves defined by $\dot \sigma_E=0$ and $\dot \sigma_D=0$, respectively, which must intercept themselves in the zero-gradient points (e.g. CY2007; see also Figure 10 in the Appendix).

For $\sigma_E$, $\sigma_D=0,\pi$, Eq. (\ref{2}) has at most five zero-gradient points: C, D, M, O and P. In the plane ($\delta,H$), the energy of each of these points as function of $\delta$ generate five curves. Fig. 9(a) shows the curves, which have been labeled with the same letter of the corresponding critical point.

The main characteristics of the resonant dynamics can be drawn from the curves plotted on the parametric plane. Very far from the resonance and for negative values of $\delta$, only the critical point M exists. In this case, the dynamics of the system is dominated by secular interactions (TW1988).

For $\delta\approx-0.005$ a bifurcation occurs in the parametric plane. For $-0.005\leq\delta\leq+0.0088$, only the curves C, D, M delimit different regions of initial conditions. The current position of the pair E-D in the parametric plane belongs to the region between C and D, and the corresponding dynamics of the system was studied in detail by CY2007. See also Figure 12 in the Appendix. Between D and M we have the co-rotation zone (see Section 3.1). For energies larger then M, no motion is possible, and we have the forbidden regions of initial parameters.

For $\delta\geq +0.0088$, a new bifurcation occurs in the plane $(\delta,H)$, and we can see the rise of the points $O$ and $P$. There are important regimes of motion in this range of $\delta$, as described in Section 3.2. For instance, the rise of the separatrix of the true libration of $\sigma_E$ around zero (L(0)), may occur for $\delta\geq +0.0088$ (see Fig. 3(f)).

The evolution of the trajectories shown in Section 3 can also be shown in parametric plane (see TW1988). In fact, as tidal evolution changes the values of $\delta$ and $H$, the system is driven to different regions of the parametric space. In Figure 11 in the Appendix we show some examples of trajectories evolving in the plane ($\delta$,$H$).

\section*{4. Discussion}

In this section, we review the scenarios of the evolution of the pair E-D through 2:1 resonance given by integrable models of resonance, and compare them with our results (Section 3). The formation of the current configuration of the pair E-D is also discussed.

In a scenario of tidal evolution, Sinclair (1972) and Henrard and Lemaitre (1983) found that the probability that the pair E-D has passed through 2:1 resonance without capture of $\sigma_D$ around $\pi$ is about $81\%$ for $e_{D0}\geq0.0032$. They considered $e_E=0$ and neglected long-period terms, that is, they used simplified models of the so-called ``Dione resonance". Remember that in adiabatic evolution through resonances considering one-degree-of-freedom models, capture into libration is certain only for initial eccentricities below a determined value $e_{max}$. Sinclair (1972) proposes that, before the resonance encounter, Dione initial eccentricity was of the order of $e_{max}\approx0.0032$, slightly larger than the current value ($\approx$0.0022), and $\sigma_D$ was circulating in retrograde direction. The passage through resonance without capture reduced $e_D$ to values near the current one, and changed the direction of circulation of $\sigma_D$.

According to our results, capture of $\sigma_D$ into true libration is a certain phenomenon whenever $e_{D0}\leq0.0035$ (see Sections 3.1 and Fig. 8). Moreover, the probability of the system to pass through resonance without capture of $\sigma_D$ is high for $e_{D0}\geq0.0035$. We conclude therefore that, in the case of the ``Dione resonance", our results are in good agreement with the classical ones.

In the case of integrable models of the ``Enceladus resonance", where $e_D=0$ and all long-period are neglected, classical results are the following: capture of $\sigma_E$ into libration about zero is guaranteed provided that $e_{E0}\leq e_{max}$. Peale (1986) and Sinclair (1972) give $e_{max}\approx0.017$ and $\approx0.019$, respectively. Considering the current data given in Table I, and using the equation of $e_{max}$ given by Peale (1986), we obtain $e_{max}\approx0.0183$. As currently $e_E\approx0.0045$, i.e., $e_E\ll e_{max}$, classical results consider that capture in libration was automatic in the past. The current small values of the free Enceladus eccentricity ($\sim 10^{-4}$; see Ferraz-Mello 1985), and amplitude of libration ($1.505^{\circ}$), would be related to tidal dissipation within the satellite.

The results shown in Section 3.2, where $0.0035\leq e_{i0}<0.017$, show that the probability of capture of $\sigma_E$ into true libration (L(0)) is high, in good agreement with classical results. However, in the case of capture into L(0), the Enceladus eccentricity could have increased to values larger than the current one. Since natural escape from this libration (case 5 in Section 3.5) is a very rare event (see Fig. 8(a)), the system could not have evolved to the current configuration.

In Section 3.3 we have studied the dynamics for $e_{E0}\gg e_{D0}$. In the cases where $e_{E0}$ is somewhat ($\geq0.017$) larger than the current one, the number of captures into L(0) is small, and the number of unsuccessful passages and escapes are large, in good agreement with the results obtained with integrable models resonance. Numerical experiments with $e_{E0}\geq0.017$ and $0.0035\leq e_{D0}\leq0.007$ have been done in order to find examples where $e_E$ and $\sigma_E$ could be reduced to the current values when the system crossed the 2:1 resonance. In spite of several tests, none of the experiments seems to be able to reproduce the current configuration of the E-D system.

Other scenarios obtained from our simulations, not discussed in classical works, are related to capture of the system into paradoxical (small-eccentricity) libration of $\sigma_E$ about zero. In Section 3.1 we have shown that, in order to get capture into paradoxical libration, we must have $e_{i0}\leq0.0015$. As we deal with general three-body problem taking into account the mutual interaction between E-D, for $e_{i0}\leq0.0015$, $\sigma_D$ usually falls into L($\pi$) regime of motion (see discussion above). This is in complete opposition to what we would expect considering the current configuration of E-D system. The capture into L($\pi$) excites the initial eccentricity of Dione, so that this libration state of $\sigma_D$ and resulting $e_D$ are not compatible with the current scenario of the E-D system.

The last case considered in Section 3 corresponds to the case where the Dione eccentricity is larger than the Enceladus one (Section 3.4). According to our results discussed above, if $e_{D0}\geq0.0035$, capture of $\sigma_D$ into $\pi$ is avoided and, if $e_{E0}\leq0.0015$, capture into small-eccentricity libration is certain. These two simultaneous events would be enough to recover the current configuration of the system. Some cases corresponding to the runs where $e_E=0$ (e.g. Figs. 7(e-h)) agree with this hypothesis: in the beginning of the integration, $\sigma_E$ is always captured into paradoxical libration while $\sigma_D$ circulates. However, the subsequent evolution of the system is marked by a complex motion, similar to the case 10 discussed in Section 3.5, where the system enters into a web of secondary resonance.

At first glance, the hypothesis $e_{D0}\gg e_{E0}$ is not plausible. Meyer and Wisdom (2008) also show some simulations of the evolution through resonance of the pair E-D where $e_{D0}\gg e_{E0}$. They also propose some additional mechanism to increase the Dione eccentricity, for instance, the passage of Dione and Mimas through the 3:1 mean-motion resonance. Alternatively, although not so probable, we can also think that some increase of the eccentricities of Dione or Enceladus could have happened during the planetary migration (Tsiganis et al. 2005; Nogueira 2008), if giant planets as Uranus and Neptune had approached very close to Saturn.

In our model, we do not reproduce the current configuration of the pair E-D in the case where $e_{D0}\gg e_{E0}$ since Enceladus eccentricity in general increases, due to capture of the system into secondary resonances, reaching values much larger than current $0.0045$. In a similar scenario, Meyer and Wisdom (2008) show an example where eccentricities decay and they can reproduce, in a limited time interval, the current configuration of the pair E-D. However, there is a main difference between the methodology applied by Meyer and Wisdom (2008) and ours. As pointed by TW1988, tidal friction acts in two different forms: the time variation of the Hamiltonian's coefficients, and the direct effects of tides on the degrees of the freedom of the problem. The latter was considered by Meyer and Wisdom (2008), but not in this work. Maybe an extension of our model by including satellites and planetary tides directly in canonical equations could lead to more exciting scenarios of evolution of the Saturn-E-D system.

\section*{5. Conclusions}
We have studied the dynamics of the pair E-D during the passage through the 2:1 mean motion resonance with a two-degrees-of-freedom planar model.
We have collected the essential physics of the general planar three-body problem (TW1988, CY2007), so that our basic dynamics is a generalization of some simplified models used by Sinclair (1972), Henrard and Lemaitre (1983) etc.

Under the action of tides, the system is slowly evolving toward the resonance, starting from several different initial conditions. The choice of these initial conditions is mostly dictated by the possible eccentricities of both satellites in the past. The simplified theory of the equilibrium eccentricity for a synchronous satellite is used to obtain a rough estimate of the range of the eccentricities.

Based on this estimative of initial eccentricities, hundreds of initial conditions (almost 1000) have been integrated considering slow tidal evolution with $Q_S=34,000$, $k_{2S}=0.34$. We summarize our main results, and the main regimes of motion associated to 2:1 mean-motion resonance are:

$\bullet$small-eccentricity (paradoxical) libration and true libration of Enceladus critical angle ($\sigma_E$), both about zero;

$\bullet$true libration of Dione critical angle ($\sigma_D$) about $\pi$;

$\bullet$libration into several secondary resonances involving linear combinations of $\sigma_E$ and $\sigma_D$.

All these regimes of motion appear in sets of runs where different initial conditions have been considered:

$\bullet$For initial eccentricities smaller than $0.0035$, $\sigma_D$ is always captured into libration about $\pi$, so that Dione eccentricity ($e_D$) increases to values larger than the current one. Moreover, in order to guarantee the capture of $\sigma_E$ into small-eccentricity libration (similar to the current one), our model requires $e_{E0}\leq0.0015$. The simultaneous occurrence of both events conflicts with the current scenario of E-D system.

$\bullet$For $0.0035\leq e_{i0}<0.017$, we have found that there is high probability of capture of $\sigma_E$ into true libration about zero and, as a consequence, $e_E$ increases to values much larger than the current one.

$\bullet$In the case where $e_{E0}\geq0.017$ and $e_{D0}\geq0.0035$, the number of captures of $\sigma_E$, $\sigma_D$ into libration is small. However, in several runs, capture into secondary resonances increases $e_E$.

$\bullet$When $e_{D0}\gg e_{E0}$, the determinant role in the evolution of the system is played by secondary resonances and complex regimes of motion involving exchange of different types of captures. In spite of the rich dynamics, no example evolving to the current E-D configuration has been verified.

We conclude that the current libration in E-D system cannot be reproduced with over-simplified models of resonance. Even with our two-degrees-of-freedom model, where several complex and chaotic motion are possible, we cannot recover in high details the current configuration of the system. More general models including more satellites and tidal dissipation in all bodies, seem to be necessary to recover the Enceladus-Dione history.
\vspace{1cm}

\textbf{ACKNOWLEDGEMENTS.} This work has been financed by FAPESP (06/58000-2, 06/61379-3). T. Yokoyama thanks FAPESP (06/04997-6), CNPQ (306276/2007-0) and FUNDUNESP.

\section*{Appendix}

\subsection*{Figure 10}
In Fig. 9(b)) we have shown levels curves of $H=H(e_E,e_D)$ for a fixed value of $\delta$. Fig. 10 shows additional examples of representative planes of initial conditions for different values of $\delta$. In Figure 10 we have also plotted, in gray and bold lines, respectively, the curves defined by $\dot \sigma_E=0$ and $\dot \sigma_D=0$. As discussed in Section 3.6, the evolution of the level curves can give important information about the dynamics of the system.

\subsection*{Figure 11}
As pointed in Section 3, as the system evolves due to the tidal dissipation, the values of $\delta$ and $H$ change, and the system passes through different regions of the parametric plane defined in Fig. 9(a). In Fig. 11, we show the variation of the energy of some trajectories in the parametric plane. Trajectories starting with eccentricities $e_{i0}<0.0015$ (e.g. dashed curves in Fig. 11(a)), in general enter in the co-rotation zone, the region located region between the curves M and D where both critical angles can librate simultaneously (see Section 3.1). For slightly larger values of eccentricity, the system evolves to the regions between D and C, far from co-rotation zone. One example is shown in Fig. 11(a,b) by gray line, and corresponds to a trajectory of case 3) discussed in Section 3.5. For initial eccentricities $e_{i0}>0.0035$, the system is, in general, captured into L(0) or into some secondary resonances. The evolution of such type of trajectories are illustrated in Fig. 11(b) by dashed curves.

\subsection*{Figure 12}
In order to help the understanding of the main regimes of motion attained by the system during its evolution on the parametric plane, we construct several dynamical maps at frozen values of $\delta$ (Fig. 12). They are based on the spectral number $N$ (see Ferraz-Mello et al. 2005 and references therein), and are built integrating the averaged equations of motion in a grid of points in the representative plane for a given $\delta$. For each point, we obtain numerically the Fourier spectrum of the solution. The spectral number $N$ is defined as the number of significant spectral peaks which amplitude is greater than $0.1\%$ of the highest peak of the variable $x_E\approx e_E\cos\sigma_E$.

In Fig. 12, we see the periodic orbits associated to the main regimes of motion of the critical angles indicated in Fig. 9(a): paradoxical libration of $\sigma_E$ around zero (denoted by PL(0)); true libration of $\sigma_E$ around zero (L(0)); true libration of $\sigma_D$ around $\pi$ (L($\pi$)); paradoxical libration of $\sigma_D$ around zero (RDIP). In order to better visualize the loci of PL(0) and L(0), we also plot in Figure 12 the loci of $\dot \sigma_E=0$ (see also Fig. 9(b)). The black regions seen in Fig. 12 correspond to the separatrix of the regime L($\pi$) associated to Dione resonance. The oval structures are associated to secondary resonances between $\sigma_E$, $\sigma_D$. Near the maximum value of energy (M) we have the co-rotation zone.

Figures 12(a,b) are similar to Fig. 8 given in CY2007, where the dynamics of the system was studied in details in these regions of the parametric plane. For larger values of $\delta$ (Figs. 12(c,d)), however, the distribution of the periodic orbits and the dynamics of the system are quite different. The maximum forced Enceladus eccentricity associated to the periodic orbit PL(0) and Dione eccentricity associated to L($\pi$), increase in the parametric plane. For larger values of $\delta$, a new chaotic zone rises between the domain of secondary resonances, which is associated to the regime of true libration of $\sigma_E$ around zero. In Figs. 12(c,d), we denote the periodic orbit associated to this regime of motion by L(0). Inspection of Fig. 12 shows that the loci of $\dot \sigma_E=0$ now represents the center of true libration L(0). See also Fig. 3(f).
\newpage
\centerline{Captions}

FIGURE 1. (a) $\dot A$ as function of $Q_S$. Vertical dashed line indicates an estimative of the minimum value of $Q_S$. Vertical full line indicates the value of $Q_S$ used in the calculation of the $\dot \delta$. (b) Equilibrium eccentricity of Enceladus ($e_{eq.}$) as function of $k_{2E}$ for several values of $Q_S$ and $Q_E$. $k_{2S}=0.34$ (see Table I).  From top to bottom: dashed curve: $Q_S=18,000$, $Q_E=100$; dashed-bold curve: $Q_S=34,000$, $Q_E=100$; full line: $Q_S=18,000$, $Q_E=20$; full-bold line: $Q_S=34,000$, $Q_E=20$. The vertical lines indicate two values of $k_{2E}$ found for Enceladus (see Table I).

\vspace{1cm}

FIGURE 2. Maximum and minimum values of the eccentricities and critical angles in fixed intervals of $\Delta \delta$, for different initial conditions. (a-d) $e_{E0}\approx e_{D0}\approx0.001$. (e-h) $e_{E0}\approx e_{D0}\approx0.002$. (i-l) $e_{E0}\approx0.0045$, $e_{D0}\approx0.0022$. In (d,h,l), the gray points show the evolution of $\Delta\varpi=\sigma_E-\sigma_D=\varpi_D-\varpi_E$.

\vspace{1cm}

FIGURE 3. (a) Enceladus section for $\delta=+0.007$ and energy in the co-rotation zone around the maximum value M (see Figs. 12 and 13 in the Appendix). (b) Dione section corresponding to (a). (c) Time evolution of $\sigma_E$, $\sigma_D$ and $\Delta\varpi$ for initial conditions in the co-rotation zone: $e_{E0}=0.0016$, $\sigma_{E0}=0$, $\lambda_{E0}=0$, $e_{D0}=0.007856$, $\sigma_{D0}=\pi$, $\lambda_{D0}=\pi$. (d) Time evolution of the pericenter of Enceladus (full circles) and Dione (open circles) of the orbit shown in (c). (e) Enceladus section for $\delta=+0.007$ and energy far from co-rotation zone. The domain of some secondary resonances are indicated by 1/2 and 1/1. (f) Enceladus section for $\delta=+0.025$.

\vspace{1cm}

FIGURE 4. Maximum and minimum values of the eccentricities and critical angles in fixed intervals of $\Delta \delta$, for different initial conditions. (a-h) Two distinct trajectories from run 8: $e_{E0}\approx e_{D0}\approx 0.007$. (a-d) $\sigma_{E0}=35.573^{\circ}$, $\sigma_{D0}=153.475^{\circ}$. (e-h) $\sigma_{E0}=1.456^{\circ}$, $\sigma_{D0}=130.431^{\circ}$. (i-l) Evolution of one trajectory from run 7: $e_{E0}\approx0.017$, $e_{D0}\approx0.0035$.

\vspace{1cm}

FIGURE 5. Maximum and minimum values of the eccentricities and critical angles in fixed intervals of $\Delta \delta$, for two distinct ((a-d) and (e-h)) trajectories from run 10, where $e_{E0}\approx0.02$, $e_{D0}\approx0.007$.

\vspace{1cm}

FIGURE 6. Maximum and minimum values of the eccentricities and critical angles in fixed intervals of $\Delta \delta$, for different initial conditions. (a-d) $e_{E0}\approx0.017$, $e_{D0}\approx0.0035$. (e-l) $e_{E0}\approx0.02$, $e_{D0}\approx0.0035$. Figs. 6(h,l) show the evolution of the angles $\sigma_E+\sigma_D$, $\sigma_E+4\sigma_D$, respectively.

\vspace{1cm}

FIGURE 7. Maximum and minimum values of the eccentricities and critical angles in fixed intervals of $\Delta \delta$, for trajectories with $e_{E0}\ll e_{D0}$. (a-d) $e_{E0}\approx0.002$, $e_{D0}\approx0.004$. (e-h) $e_{E0}\approx0$, $e_{D0}\approx0.008$. (i,j) The same as (a,c), respectively, where $e_{E0}\approx0$, $e_{D0}\approx0.002$. (k,l) Same as (a,c), respectively, where $e_{E0}\approx0$, $e_{D0}\approx0.004$. Figs. 7(d,h) show the evolution of the angles $2\sigma_E+\sigma_D$, $\sigma_E+\sigma_D$, respectively.

\vspace{1cm}

FIGURE 8(a,b). Number of occurrences of the regimes of motion 1-11) listed in Section 3.5 as function of the \emph{initial} eccentricities. The final values of $\delta$ attained in different runs are shown on the top of the bars. (a) $e_{E0}\sim e_{D0}$. $\delta_{final}\approx0.0324$ in the interval $e_{i0}=0$ and $e_{i0}=0.003$. (b) $e_{E0}\gg e_{D0}$. (c) $e_{E0}\ll e_{D0}$.

\vspace{1cm}

FIGURE 8(c). Continued.

\vspace{1cm}

FIGURE 9. (a) The paths of zero-gradient points M, D, C, P and O in the parametric plane ($\delta,H$). The localization of the main regimes of motion are indicated by symbols; black triangle: retrograde or direct circulation of $\sigma_E$, $\sigma_D$; big circle: paradoxical libration of $\sigma_E$, $\sigma_D$ around zero; $\bullet$: true libration of $\sigma_E$ around zero; black square: true libration of $\sigma_D$ around $\pi$; square: paradoxical libration of $\sigma_E$ around $\pi$; secondary resonances. (b) Representative plane ($e_E,e_{D}$) of initial conditions for $\delta=+0.018$. Each curve is a level curve of the Hamiltonian (\ref{2}), for symmetric configurations. Positive and negative values of eccentricity correspond to $\sigma_i=0$ and $\pi$, respectively. The gray points and bold line gives the loci of $\dot \sigma_E=0$ and $\dot \sigma_D=0$, respectively.

\vspace{1cm}

FIGURE 10. Representative planes ($e_E,e_D$) of initial conditions for different values of $\delta$ parameter. Each curve is a level curve of the Hamiltonian (2), for symmetric configurations. Positive and negative values of eccentricity correspond to $\sigma_i=0$ and $\pi$, respectively. The gray and black-bold lines are the loci of $\dot \sigma_E=0$ and $\dot \sigma_D=0$, respectively. (a) $\delta=-0.014$. (b) $\delta=+0.005$. (c) $\delta=+0.0088$. (d) $\delta=+0.018$.

\vspace{1cm}

FIGURE 11. The paths of zero-gradient points M, D, C, P and O in the ($\delta,H$) plane. (a) Typical trajectories taken from runs with eccentricities near zero: Run 1 ($e_{E0}\approx e_{D0}\approx0.001$, dashed lines), Run 2 ($e_{E0}\approx e_{D0}\approx0.002$, gray line). (b) Several trajectories from Run 8 ($e_{E0}\approx e_{D0}\approx0.007$, dashed lines), and Run 2.

\vspace{1cm}

FIGURE 12. Dynamical maps showing the spectral number $N$, as defined in the text, for several values of $\delta$. (a) $\delta=0$. (b) $\delta=+0.007$. (c) $\delta=+0.01$. (d) $\delta=+0.018$. The bold-gray lines show the loci of $\dot \sigma_E=0$ obtained analytically and match the loci of the periodic orbits of PL(0) and L(0).

\end{document}